# Local and Global Screening Properties of Graphene Revealed through Landau Level Spectroscopy.

Chih-Pin Lu[1], Martin Rodriguez-Vega[2], Guohong Li[1], Adina Luican-Mayer[1], K. Watanabe[3], T. Taniguchi[3], Enrico Rossi[2] and Eva Y. Andrei[1]

[1] Department of Physics and Astronomy, Rutgers University, Piscataway, New Jersey 08855, USA

[2] Department of Physics, College of William and Mary, Williamsburg, VA 23187, USA

[3] Advanced Materials Laboratory, National Institute for Materials Science, 1-1 Namiki, Tsukuba 305-0044, Japan

**Abstract**

One-atom thick crystalline layers and their vertical heterostructures carry the promise of designer electronic materials that are unattainable by standard growth techniques. In order to realize their potential it is necessary to isolate them from environmental disturbances in particular those introduced by the substrate. But finding and characterizing suitable substrates, and minimizing the random potential fluctuations they introduce, has been a persistent challenge in this emerging field. Here we show that Landau-level (LL) spectroscopy is exquisitely sensitive to potential fluctuations on both local and global length scales. Harnessing this technique we demonstrate that the insertion of an intermediate graphene layer provides superior screening of substrate induced disturbances, more than doubling the electronic mean free path. Furthermore, we find that the proximity of hBN acts as a nano-scale "vacuum cleaner", dramatically suppressing the global potential fluctuations. This makes it possible to fabricate high quality devices on standard $SiO_2$ substrates.

TOC Graphic:

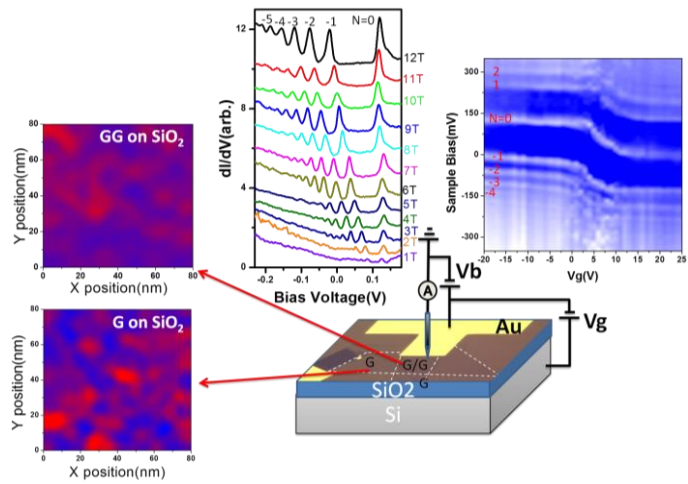

The recent realization of one-atom thick layers and the fabrication of layered Van der Waals heterostructures revealed fascinating physical phenomena and novel devices based on interlayer interactions[1-9]. Inherent to the 2D structure of these layers is an extreme vulnerability to disturbances introduced by the substrate[10-13]. Substrate interference can be eliminated by suspending the sample, an approach that led to the observation of ballistic transport[14-20] and the fractional quantum Hall effect in graphene[16-18], but its application is limited to small micron-size samples at relatively low doping. Another approach is to use atomically smooth metallic substrates[21-23] or graphite[24-28] which screen the random potential. But these substrates short-circuit the 2D channel and prevent tuning the carrier density by gating, rendering them unsuitable for device applications. Among insulating substrates atomically flat hBN [29-31] and $MoS_2$[7] have recently emerged as promising alternatives to $SiO_2$ substrates.

Using scanning tunneling spectroscopy (STM), spectroscopy (STS) and numerical simulations we demonstrate that inserting a graphene buffer layer between the 2D sample (in this case also graphene) and the insulating substrate, screens the random potential fluctuations without compromising the gating capability. We characterize the effect of local potential fluctuations on a local scale using LL spectroscopy, a technique which gives direct access to the onset of well-defined cyclotron orbits, to the quasiparticle life time and to the mean free path. Furthermore, we show that gate dependent LL spectroscopy allows to quantify the global potential fluctuations. Using this technique we find that the presence of a graphene buffer layer and the proximity of an hBN flake result in a dramatic suppression of the global potential fluctuations.

Devices were fabricated from exfoliated graphene flakes and transferred onto the surface of a 300 nm chlorinated $SiO_2$ layer capping a highly n-doped Si substrate, which served as a back-gate. To ensure decoupling between the top and bottom graphene layers and to avoid interference from Van Hove singularities,[2, 3, 9] the layers were deposited with a large twist angle between them. Standard e-beam lithography followed by electron-beam evaporation at base pressure of 2x10⁻⁷ Torr was employed to deposit the Ti/Au (2 nm/60 nm) pads for guiding the STM tip to the sample[32]. The devices were baked for 3hrs in forming gas at 250 $^0$C prior to mounting into the cryostat. STM and STS measurements were performed at 4 K in a home-built STM using Pt-Ir tips that were mechanically cut from polycrystalline wire. STM images are recorded in constant current mode with the bias voltage, $V_b$, applied between the sample and grounded tip. Differential conductance (dI/dV) spectra which are proportional to the local DOS, were obtained with a lock-in technique at modulation frequency 440 Hz with fixed tip to sample distance.

The schematic measurement set up is shown in Figure 1(a). Figure 1(b) shows STM topography of a single layer (GSiO$_2$) and an adjacent double layer (GGSiO$_2$). The step height across the boundary between the two regions, ~0.7 nm, is significantly larger than for Bernal stacked graphite (0.34 nm)[33], suggesting that the top and bottom layers are electronically decoupled. Side by side topography images on residue-free regions of the GSiO$_2$ and GGSiO$_2$ regions, Figure 1(c), show that they have the same average height corrugation of ~ 0.9 nm [Figure 1(d)]. The absence of a moiré pattern in the GGSiO$_2$ sample suggests weak interlayer coupling, consistent with their large separation.

Figure 2(a) shows the gate voltage, $V_g$, dependence of the dI/dV spectra, which are proportional to the local DOS, for the GGSiO$_2$ sample (see Fig. S1 for GSiO$_2$). In Figure 2(b) we plot the Dirac point (DP) energy, $E_D$, obtained from panel (a) as a function of $V_g$. Fitting to the expression expected for single layer graphene,[34] $E_D = \hbar v_F \sqrt{\frac{1}{2}\pi\alpha|V_g - V_0|}$, we obtain the Fermi velocity $v_F = (1.02 \pm 0.04) \times 10^6$ m/s consistent with the accepted value for graphene on SiO$_2$. Here $\hbar$ is the reduced Planck constant and $\alpha = 7\times 10^{10}$ cm$^{-2}$ V$^{-1}$ is the charging capacitance per layer, per unit area and unit charge. The offset, $V_0 = 22.5 \pm 0.5$ V, indicates unintentional hole-doping with carrier density, n ~ $8\times 10^{11}$ cm$^{-2}$, per layer.

The substrate-induced random potential produces electron-hole puddles observed as DOS fluctuations in the maps shown in Figures 2(c) and 2(d) for GSiO$_2$ and GGSiO$_2$ respectively. We note that the fluctuation amplitude is reduced in the double layer compared to the single layer and the average puddle size increased from ~ ($13 \pm 3$) nm in the single layer to ($21 \pm 3$)nm in the double layer (Figure S2(a) and (b) ), reflecting an almost doubling in screening length afforded by introducing the bottom graphene layer.

To further understand screening in this system we carried out numerical simulations. In graphene, unlike the case of materials with parabolic bands, the disorder potential created by trapped charges retains its long-range nature[35-37]. This, together with the non-linear nature of the screening in graphene[38] poses significant challenges to theoretical treatments. Previous work has shown that the Thomas-Fermi-Dirac theory (TFDT)[38, 39] provides a computationally feasible approach to modeling this problem (SI3). Starting from the random distribution of charge impurities shown in Figure S3(a) and (b) we used TFDT to numerically illustrate the screening effect of a single graphene layer, Figure 2(e), and to demonstrate the shielding effect of adding a second layer, Figure 2(f). Similarly to the experimental results, we find that the double-layer experiences a substantial reduction in the potential fluctuations compared to the single layer.

LL spectroscopy makes it possible to quantify the screening effect by providing access to the scattering length, the quasiparticle lifetime, and the potential fluctuations [27]. In the presence of a magnetic field, B, normal to the layer the spectrum breaks up

into a sequence of LLs: [34]

$$E_N = E_D \pm \frac{\hbar v_F}{l_B}\sqrt{2|N|} \quad N=0, \pm 1, \pm 2, \pm 3, ... \quad (1)$$

where $l_B \approx \left(\frac{\hbar}{eB}\right)^{1/2}$ is the magnetic length. The LLs become observable when their characteristic energy scale, $\hbar v_F/l_B$ exceeds, the line-width $\Delta E = \Delta E_{lw} + \Delta E_D$, where $\Delta E_{lw}$ is the intrinsic linewidth and $\Delta E_D$ is the potential fluctuation amplitude across a cyclotron orbit. This criterion implies: $l_B < v_F \tau = l$, where $\tau \approx \hbar/\Delta E$ is the quasiparticle lifetime and $l$ is the mean-free-path, which in 2D is also the screening length[40]. In other words LL become observable when the cyclotron orbit can "fit" within a uniform puddle of charge. This defines an onset field, $B_o \sim \left(\frac{\Delta E}{v_F}\right)^2 \frac{1}{e\hbar}$, above which the first LL becomes observable. For lower fields, $B < B_o$, the behavior is dominated by scattering from the random potential. Thus $B_o$ measures the local random potential fluctuations, providing a gauge of substrate quality.

The evolution of LLs with field is shown in Figures 3(a) and (b). Once the LLs are observed their energy follows the dependence expected for single layer graphene, $E_N \propto \sqrt{|N|B}$, as shown in Figure S4(a). Fitting the data to equation (1) we find $v_F = (1.10 \pm 0.02) \times 10^6$ m/s and $(1.12 \pm 0.01) \times 10^6$ m/s for the single and double layer respectively. Both values are consistent with that obtained from the zero-field gate dependence For GGSiO$_2$ the peaks are sharper and their onset is earlier than in GSiO$_2$ indicating a more homogeneous charge distribution and a longer quasiparticle lifetime. We find $B_o \sim 3.5$T and $\sim 0.5$T corresponding to $\Delta E \sim 46$ meV and $\sim 18$ meV for the single and double layer, respectively. Gaussian fits of the N=0 LL (Fig. S4), gives comparable values: $\Delta E \sim 42 meV$ and 18 meV for the single a double layer respectively. These results correspond to more than doubling the carrier lifetimes, from $\tau \approx 15$ fs for a single layer, to $\tau \approx 35$ fs for the double layer and to a similar increase in the mean-free-path from $l \sim 15$ nm to $\sim 35$ nm, demonstrating that the use of the buffer graphene layer significantly reduces the local potential fluctuations. Interestingly these values of the mean-free-path are comparable to the average puddle size obtained in Fig.2(c) and 2(d), supporting the idea that LLs become observable when the cyclotron orbit "fits" inside a charge puddle. Using TFDT simulations with the same parameters as those in Figs. 2(e), and 2(f), we find that the disorder averaged values for $\Delta E_D$, 31 meV and 14 meV for single and double layer respectively, agree with the experimental values (SI3 Figure (c) and (d)). Comparing to the results obtained for graphene on graphite[25-28] the LL line-widths measured here are broader and they are independent of energy indicating that the scattering, although reduced, is still extrinsic.

In Figure 3(c) we illustrate the effect of an hBN flake placed close to the double layer (Figure S5(a)). The onset field, ~0.5 T, and the linewidth, $\Delta E_l \sim 17\ meV$, are not very different than with the hBN flake. But as we show next, even though it is not part of the graphene substrate, the mere proximity of the hBN suppresses the global potential fluctuations. This is consistent with earlier reports of self-cleansing at the graphene-hBN interface which is believed to segregate contaminants leaving the rest of the interface atomically clean[8].

We have seen that $B_o$ reflects the scale of the local potential fluctuations. Now we show that the LL spectra also provide access to the global potential fluctuations across the entire sample. In the absence of fluctuations the gate dependence of the LLs produces a staircase pattern consisting of a sequence of equidistant plateaus separated by sharp jumps[41, 42]. The plateaus reflect the pinning of $E_F$ within a LL as it is being filled, and their width, $\Delta V_g = \frac{8}{\alpha}\frac{B}{\phi_0}$, corresponds to the gate voltage needed to populate one LL in each layer [42]. Here $\phi_0 = 4.14 \times 10^{-15}$ Tm$^2$ is the fundamental unit of flux and 8 reflects the degeneracy due to spin, valley and layers. Once the N'th LL is filled, $E_F$ jumps to the next LL producing the sharp step. In STS measurements $E_F$ defines the energy origin and therefore it is $E_D$ and the LLs that appear to be shifting rather than $E_F$. A random potential smears out the staircase structure because $E_D$ (and the entire LL sequence with it) fluctuates across the sample following the potential variations. It is important to note that $V_g$, being controlled by the gate electrode, is depositing charge across the entire sample and not only at the position of the STM tip. Therefore, as $V_g$ is swept the first electron to populate a given LL will occupy a state localized near the global minimum of the random potential while the last electron will find a state near the global maximum causing the LL energy to trace out the global random potential fluctuation, $\Delta E_{DG}$. As a result the plateaus acquire a slope which tracks the global random potential distribution across the entire sample.

In Figures 4(a), (b) and (c) we show the gating effect on the LL spectra for GSiO$_2$, GGSiO$_2$, and GGSiO$_2$ near hBN, respectively. In the case of GSiO$_2$ the absence of an observable staircase structure signifies that $\Delta E_{DG}$ exceeds the LL spacing at 10T ~ 115 meV. This is more than three-fold larger than the local fluctuation amplitude obtained from the onset field. For the GGSiO$_2$ sample the staircase becomes discernible and from the plateau slope we find $\Delta E_{DG}$ ~50 meV. Thus, adding a second graphene layer strongly suppresses the substrate induced disorder on both local and global scales. Most remarkably when the sample is close to an hBN flake the global potential fluctuations are almost completely suppressed as shown in Figure 4(c) and also S5(b) and (c). Now the plateaus are much flatter with an estimated slope of ~ 11 meV, corresponding to a reduction of $\Delta E_{DG}$ below the LL linewidth, directly demonstrating the efficacy of the self-cleansing phenomenon of hBN.

In summary, LL spectroscopy is a remarkably sensitive diagnostic tool for characterizing the local and global electronic properties of 2D layers. Using this technique we demonstrate that substrate intrusion is substantially reduced by introducing a graphene buffer layer between the substrate and an atomic layer of interest.

Moreover, as a result of a still poorly understood self-cleansing phenomenon, the proximity of hBN suppresses the global potential fluctuations. Although these experiments were carried out on graphene we expect similar shielding and cleansing effects for other atomically thin layers.

Funding was provided by DOE-FG02-99ER45742 (characterization), NSF DMR 1207108 (STM STS), EFRI-2DARE 1433307 (fabrication). We thank Ivan Skachko and Jinhai Mao for useful discussions.

.

FIGURE CAPTIONS:

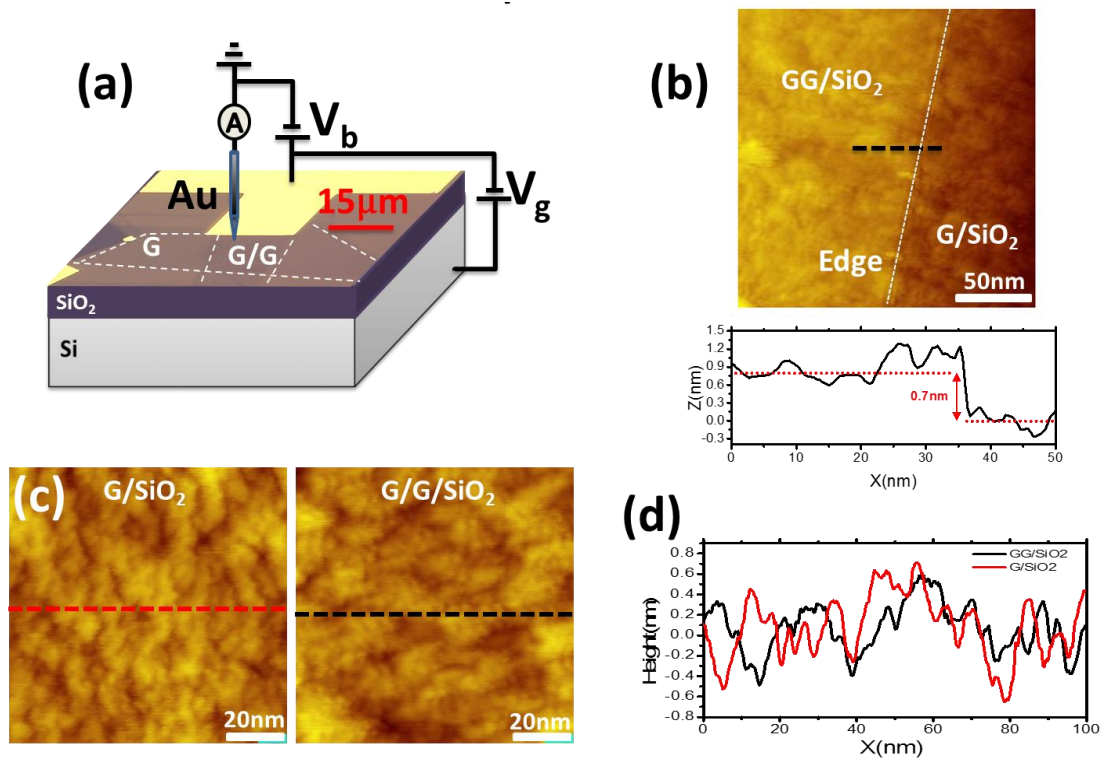

Figure 1. (a) Optical micrograph of the GSiO$_2$ and GGSiO$_2$ samples and Ti/Au electrode, shown with the schematic STM setup. (b) (top)Constant current STM topography map of the boundary between GGSiO$_2$ and GSiO$_2$; (bottom) line cut along the dashed line crossing the boundary shows a step height of 0.7nm. Tunneling parameters $I_{set}$ = 20 pA and $V_b$ = 0.7 V. (c) Constant current STM topographs of GSiO$_2$ and GGSiO$_2$. Tunneling parameters $I_{set}$ = 20 pA and $V_b$ = 0.4 V. (d) Height profiles along the dashed lines in (c).

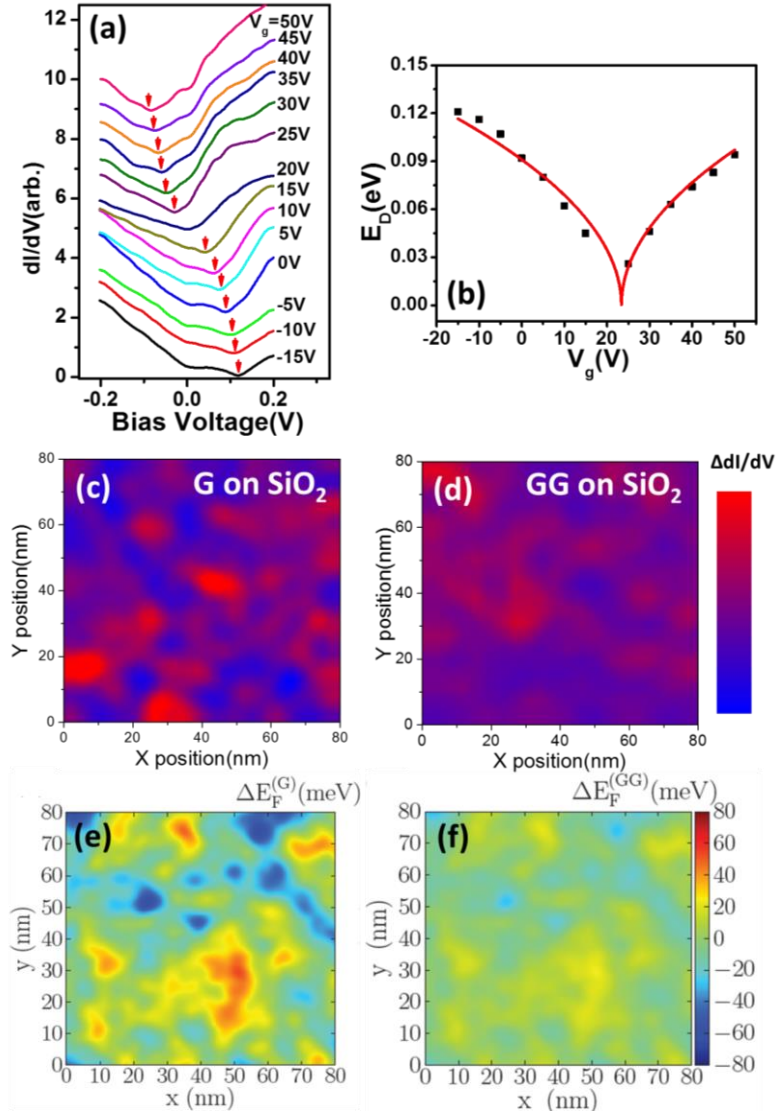

Figure 2. (a) Gate voltage dependence of dI/dV spectra on GGSiO$_2$ sample. Curves are vertically displaced for clarity. Red arrows indicate the conductance minimum which is identified with E$_D$. Tunneling parameters: I$_{set}$ = 20 pA , V$_b$ = 0.3 V, modulation voltage 5 mV.  (b) Gate voltage dependence of measured E$_D$ (squares) together with the fit (solid line) discussed in the text.  (c) and (d) dI/dV maps at V$_b$ = 0.3 V reveal the electron (red) hole (blue) puddles resulting from doping inhomogeneity. Maps were over the same area as in Figures 1(c). The color scale which is proportional to the deviation of dI/dV from the mean value across the map, is a direct representation of the local fluctuations of E$_D$. (e) and (f) Simulated map illustrating the spatial fluctuations of E$_D$ for a single disorder realization (shown in S3(a) and (b)) for a graphene single layer (e) and double layer (f). Simulation parameters: impurity density n$_{imp}$ = 5 × 10$^{11}$cm$^{-2}$, carrier density <n> = 1 ×10$^{12}$cm$^{-2}$, distance above substrate 3 nm, interlayer distance 0.7nm.

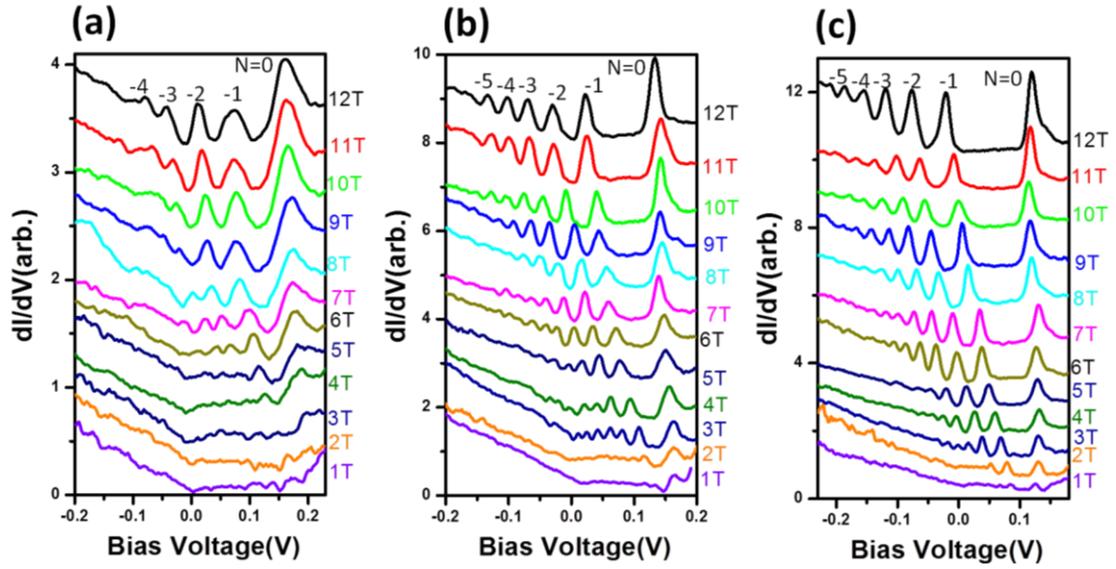

Figure 3. Field dependence of LL spectra. (a) GSiO$_2$, V$_g$ = 10 V; (b) GGSiO$_2$, V$_g$ = -15 V; (c) GGSiO$_2$ in the vicinity of hBN, V$_g$ = -10 V. All curves are offset vertically for clarity. The LL indexes, N = 0,-1,-2,-3,... are marked. STS parameters: I$_{set}$ = 20 pA, sample bias V$_b$ = 0.3 V and modulation voltage 2 mV.

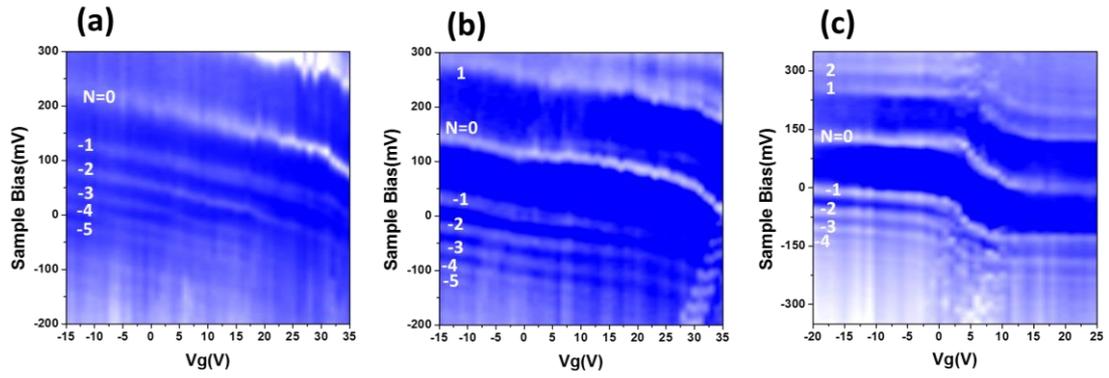

Figure 4. Gate voltage maps of LLs at 10 T. Each vertical line represents a LL spectrum at a particular $V_g$. (a)GSiO$_2$, (b)GGSiO$_2$ and (c)GGSiO$_2$ in the vicinity of hBN. The LL indexes, N = 0, ± 1, ±2…. are marked. STS parameters: $I_{set}$ = 20 pA, sample bias $V_b$ = 0.3 V and modulation voltage 5 mV.